\documentclass[letterpaper,preprint,byrevtex,showpacs,groupedaddress,pre]{revtex4-1}

\DeclareSymbolFont{lettersA}{U}{pxmia}{m}{it}
\DeclareMathAlphabet{\mathsfsl}{OT1}{cmss}{m}{sl}

\SetSymbolFont{lettersA}{bold}{U}{pxmia}{bx}{it}
\DeclareFontSubstitution{U}{pxmia}{m}{it}
\DeclareSymbolFontAlphabet{\mathfrak}{lettersA}
\DeclareMathSymbol{\piup}{\mathord}{lettersA}{"19}
\DeclareMathSymbol{\iTheta}{\mathalpha}{letters}{2}

\usepackage{amsmath}
\usepackage{amssymb}
\usepackage{multirow}
\usepackage{graphicx}
\usepackage{epstopdf}
\usepackage[hang,scriptsize]{subfigure}
\usepackage{bbm}
\usepackage{mathrsfs}
\usepackage{xcolor}
\usepackage[colorlinks=true,pdfstartview=FitV,linkcolor=blue,citecolor=blue,urlcolor=blue]{hyperref}

\makeatletter

\newcommand{\Rmnum}[1]{\expandafter\@slowromancap\romannumeral #1@}
\makeatother

\renewcommand{\vec}[1]{\boldsymbol{#1}}

\newcommand{\ii}{\mathrm{i}}
\newcommand{\diff}{\mathrm{d}}

\begin{document}

\title{Realizing broadband electromagnetic transparency with a graded-permittivity sphere}

\author{Lei~Sun}
\affiliation{Department of Mechanical and Aerospace Engineering, \\
    Missouri University of Science and Technology, \\
    Rolla, Missouri 65409, USA}

\author{Jie~Gao}
\email[Electronic address: ]{gaojie@mst.edu}
\affiliation{Department of Mechanical and Aerospace Engineering, \\
    Missouri University of Science and Technology, \\
    Rolla, Missouri 65409, USA}

\author{Xiaodong~Yang}
\email[Electronic address: ]{yangxia@mst.edu}
\affiliation{Department of Mechanical and Aerospace Engineering, \\
    Missouri University of Science and Technology, \\
    Rolla, Missouri 65409, USA}

\begin{abstract}
Broadband electromagnetic transparency phenomenon is realized with
a well-designed graded-permittivity sphere, which has an extremely
low scattering cross section over a wide frequency range, based on
the generalized Mie scattering theory and numerical simulation in
full-wave condition.
The dynamic polarization cancellation is revealed by studying the
variation of the polarization with respect to the frequency.
Furthermore, a properly-designed multi-shell sphere is also proposed
and examined in order to reduce the rigorous conditions for realizing
the broadband transparency in experiments.
\end{abstract}


\maketitle

\section{Introduction}

Metamaterials, the artificial media with unit cells smaller than the wavelength,
possess unique electromagnetic properties unavailable in nature.
Besides the well-known property of negative refraction \cite{Veselago1968SPU,Shelby2001Sci}
that leads to imaging beyond the diffraction limit \cite{Pendry2000Sci,Liu2007Sci},
electromagnetic transparency is one intriguing capability of metamaterials,
which has been extensively studied in physics and engineering communities
since it has significant impact in various fields, including optics,
medicine, biology, and nanotechnology.
In order to achieve the electromagnetic transparency phenomenon, several different
mechanisms have been proposed and examined in the past years.
Transformation optics \cite{Leonhardt2006Sci,Pendry2006Sci}
suggests covering the object with a sophisticated design of heterogeneous and
anisotropic metamaterial shell that can bend the incident electromagnetic wave
around the cloaked object.
Although the transformation optics has been well developed
in theory \cite{Leonhardt2009PO,Lenohardt2009Sci,Tyc2010IEEE}
and demonstrated in experiments \cite{Schurig2006Sci,Valentine2009NM},
it is still difficult to implement due to the approximated ideal parameters,
high material losses, and the single operating frequency \cite{Chen2007PRL}.
On the other hand, the scattering cancellation
\cite{Kerker1975JOSA,Chew1976JOSA,Alu2005PRE,Alu2007OE1},
also known as the plasmonic cloaking, suggests a properly-designed material
shell to reduce the scattering of the covered object near the plasmon resonance
frequency in order to make the object-shell system transparent.
The fundamental principle behind this mechanism is that the polarization of
the shell is out of phase with respect to the polarization of the covered
object, thus the total polarization of the object-shell system is close to
zero to the probing electromagnetic wave.
Although the scattering cancellation for transparency is
robust \cite{Alu2007OE2,Alu2008JOA}
and can be achieved in experiments \cite{Edwards2009PRL,Muhlig2013SR},
the operating frequency is still limited to the plasmon resonance frequency
of the system.
Furthermore, the electromagnetic transparency can also be accomplished by applying
the active cloaking \cite{Miller2006OE,Vasquez2009PRL,Vasquez2009OE,
Ma2013PRL,Selvanayagam2013PRX},
in which the covered object is surrounded by active sources that can tailor the
scattering automatically with respect to the properties of the probing electromagnetic
wave.
However, the active cloaking needs the whole prior knowledge about the probing
electromagnetic wave, such as frequency and phase, in order to determine the
response of active sources.
Meanwhile, the operating frequency range of the active cloaking approach is also
strongly limited due to the dimensions of the active sources.

In previous work, the electromagnetic scattering from a graded-permittivity
sphere is theoretically studied based on the generalized Mie scattering theory
and the broadband electromagnetic transparency is revealed under the quasi-static
condition \cite{Sun2011JOSAB}.
To further explore the physical mechanism dominating such unique phenomenon,
the broadband electromagnetic transparency is realized and studied in the
rigorous full-wave condition based on the numerical simulations.
A dynamic polarization cancellation associated to the frequency variation of
the probing electromagnetic wave is achieved in a well-designed sphere with
graded-permittivity along the radial direction.
Since scattering cross section is a general physical quantity to describe the
degree of the electromagnetic transparency \cite{Monticone2013PRX},
such graded-permittivity sphere is designed to have an extremely low scattering
cross section over a wide frequency range in order to obtain the broadband
electromagnetic transparency.
The dynamic polarization cancellation with respect to the variation of the
frequency is clearly revealed from the polarization analysis of the
graded-permittivity sphere.
Furthermore, a multi-shell sphere with homogeneous and isotropic permittivity in each
shell corresponding to the graded-permittivity sphere is also proposed
and examined in order to reduce the rigorous conditions for realizing
the broadband transparency in experiments.

\section{Theoretical Analysis}

Consider a dilute metal-dielectric composite, in which the permittivity of
the metallic solute is represented by a simple lossless Drude model
$\varepsilon_{m}=\varepsilon_{\infty}-\omega_{p}^{2}/\omega^{2}$
and the permittivity of the dielectric solvent is denoted as $\varepsilon_{d}$,
the effective permittivity of the composite can be approximately determined
according to the simple mixing rule as
$\varepsilon_{\mathrm{eff}}=f_{m}\varepsilon_{m}+(1-f_{m})\varepsilon_{d}
=f_{m}(\varepsilon_{\infty}-\varepsilon_{d})+\varepsilon_{d}-f_{m}\omega_{p}^{2}/\omega^{2}$,
where $f_{m}$ is the filling ratio of the metallic solute.
The expression of the effective permittivity implies that if the filling ratio
of the metallic solute can continuously vary along a certain direction in the
metal-dielectric composite, the mixture can possess a graded effective permittivity
in order to achieve broadband electromagnetic properties.
Based on this idea, the scattering of a plane electromagnetic wave
from an isotropic nonmagnetic sphere with the radius $r_{0}$ and a graded
electric permittivity varying along the radial direction is studied,
as depicted in Fig.~1(a).
The graded electric permittivity tensor is described as
$\vec{\varepsilon}(\omega,r)=\varepsilon_{0}\varepsilon_{g}(\omega,r)\vec{I}$,
in which $\varepsilon_{0}$
is the electric permittivity of the isotropic background medium (vacuum)
and the relative permittivity $\varepsilon_{g}(\omega,r)$ is characterized
by a simple lossless graded Drude model
\begin{equation}
\label{eq:grade-drude}
    \varepsilon_{g}(\omega,r) = 1 - \omega_{p}^{2}/\omega^{2}
        \left[ C_{0} - C_{1} \left( r/r_{0} \right)^{2} \right],
\end{equation}
and $\vec{I}=\vec{e}_{r}\vec{e}_{r}+\vec{e}_{\theta}\vec{e}_{\theta}+\vec{e}_{\phi}\vec{e}_{\phi}$
is the unit tensor in the spherical coordinates.
In addition, parameters $C_{0}$ and $C_{1}$ are constants (graded parameters)
controlling the variation of the electric permittivity, which are optimized as
$C_{0}=0.5$ and $C_{1}=0.9$ in order to realize the broadband electromagnetic
transparency according to the previous theoretical analysis \cite{Sun2011JOSAB}.
The variation of the corresponding graded electric permittivity is
illustrated in Fig.~1(b).
It is clear that over the wide frequency range from zero to the plasma
frequency $\omega_{p}$ the graded-permittivity sphere possesses both
negative and positive permittivity along the radial direction, which
leads to polarizations with opposite directions in the graded-permittivity
sphere with respect to the incident electromagnetic wave.
The opposite-direction polarizations can cancel each other out and result
in a small dipole moment and therefore a small scattering cross section
over a wide frequency range, so that the broadband electromagnetic
transparency can be realized.
In addition, it is worth mentioning that the radius of the graded-permittivity
sphere is set as $r_{0}=\lambda_{p}$ associated with the plasma frequency $\omega_{p}$.
Furthermore, the cancellation mechanism of the opposite-direction
polarizations is not related to any resonance phenomena in the system,
which means it is not strongly sensitive to the variation of the material
loss \cite{Alu2005PRE}, thus the material loss is ignored in order to emphasize
the polarization cancellation phenomena in the graded-permittivity sphere.

Theoretically, the scattering cross section can be calculated via the
generalized Mie scattering theory by introducing two scale potential
functions $u(r,\theta,\phi)$ and $v(r,\theta,\phi)$ and expressing the
electromagnetic field as \cite{Sun2011JOSAB}
\begin{eqnarray}
\label{eq:potential_field_E}
    \vec{E} &= \frac{1}{\ii\omega\vec{\varepsilon}(\omega,r)}
        \vec{\nabla}\times\vec{\nabla}\times(\vec{r}u) + \vec{\nabla}\times(\vec{r}v), \\
\label{eq:potential_field_H}
    \vec{H} &= \frac{1}{\ii\omega\vec{\mu}(\omega,r)}
        \vec{\nabla}\times\vec{\nabla}\times(\vec{r}v) + \vec{\nabla}\times(\vec{r}u).
\end{eqnarray}
Subsequently, regarding the incident electromagnetic plane wave
$\vec{E}_{\mathrm{in}}=\vec{e}_{x}e^{\ii kr\cos\theta}$,
$\vec{H}_{\mathrm{in}}=(\vec{e}_{y}/Z)e^{\ii kr\cos\theta}$,
where $k = \omega\sqrt{\varepsilon_{0}\mu_{0}}$
and $Z=\sqrt{\mu_{0}/\varepsilon_{0}}$,
the scale potential functions for the scattering field read \cite{Sun2011JOSAB}
\begin{eqnarray}
    u_{\mathrm{sca}}(r,\theta,\phi) = &-\frac{\omega \varepsilon_{0}}{k}
        \sum_{\ell=0}^{\infty}\ii^{\ell+1}\frac{2\ell+1}{\ell(\ell+1)}
        A_{u}^{(\ell)}h_{\ell}^{(1)}(kr)P_{\ell}^{(1)}(\cos\theta)\cos\phi, \\
    v_{\mathrm{sca}}(r,\theta,\phi) = &-\frac{\omega \mu_{0}}{kZ}
        \sum_{\ell=0}^{\infty}\ii^{\ell+1}\frac{2\ell+1}{\ell(\ell+1)}
        A_{v}^{(\ell)}h_{\ell}^{(1)}(kr)P_{\ell}^{(1)}(\cos\theta)\sin\phi,
\end{eqnarray}
with respect to the first kind spherical Hankel function $h_{\ell}^{(1)}(kr)$
and the associated Legendre polynomials $P_{\ell}^{(1)}(\cos\theta)$.
On the other hand, the scale potential functions for the electromagnetic field
in the graded-permittivity sphere read \cite{Sun2011JOSAB}
\begin{eqnarray}
\label{eq:potential_in_u}
    u_{\mathrm{sp}}(r,\theta,\phi) = &-\frac{\omega \varepsilon_{0}}{k}
        \sum_{\ell=0}^{\infty}\ii^{\ell+1}\frac{2\ell+1}{\ell(\ell+1)}
        B_{u}^{(\ell)}f_{u}(r)P_{\ell}^{(1)}(\cos\theta)\cos\phi, \\
\label{eq:potential_in_v}
    v_{\mathrm{sp}}(r,\theta,\phi) = &-\frac{\omega \mu_{0}}{kZ}
        \sum_{\ell=0}^{\infty}\ii^{\ell+1}\frac{2\ell+1}{\ell(\ell+1)}
        B_{v}^{(\ell)}f_{v}(r)P_{\ell}^{(1)}(\cos\theta)\sin\phi,
\end{eqnarray}
in terms of the function $f_{u}(r)$ and $f_{v}(r)$ that are related to the confluent
Heun function and the confluent hypergeometric function, respectively.
Here the coefficients $A_{u}^{(\ell)}$ and $A_{v}^{(\ell)}$ (the scattering coefficients),
as well as $B_{u}^{(\ell)}$ and $B_{v}^{(\ell)}$ are all determined by the boundary
conditions at the surface of the graded-permittivity sphere.
In addition, the scattering coefficients are directly associated
to the scattering cross section
\begin{equation}
\label{eq:sca}
    C_{\mathrm{sca}}=\frac{2\pi}{k^2}\sum_{\ell=1}^{\infty}
        (2\ell+1)\left( \left|A_{u}^{(\ell)}\right|^2
        + \left|A_{v}^{(\ell)}\right|^2 \right)
\end{equation}
as a summation of an infinite series.

\section{Numerical Analysis}

In theoretical analysis, it is impossible to take all terms into account in the
summation of Eq.~(\ref{eq:sca}), thus a properly determined cut-off term of the
summation is very critical.
Therefore, numerical simulation is performed to calculate the scattering cross
section of the graded-permittivity sphere.
The efficiency factor of the scattering cross
section $\sigma_{\mathrm{sca}}$ (scattering efficiency for short),
which is defined as the ratio of the scattering cross section and
the cross section of the scatter itself,
is examined based on the finite-element method (FEM).
Figure~2(a) presents the variation of the scattering efficiency for the
graded-permittivity sphere.
For comparison, the scattering efficiency for a normal metallic-like sphere
[$C_{1}=0$ in Eq.~(\ref{eq:grade-drude})] is also calculated and plotted.
It is shown that the scattering efficiency for the graded-permittivity sphere
is extremely small compared with that of the normal sphere over the wide
frequency range due to the cancellation of the opposite-direction
polarizations of the graded-permittivity sphere.
Especially, the cancellation of the opposite-direction polarizations also
reduces the localized surface plasmon resonance on the graded-permittivity sphere,
which leads to an extremely small scattering efficiency at the resonance
frequency for the graded-permittivity sphere.
However, strong localized surface plasmon resonance can be observed in the
normal sphere.
In order to reveal the polarization cancellation mechanism, the dipole moment
of the graded-permittivity sphere is studied, which is defined as the integration
of the polarization $\vec{P}$ [determined by the electric field in the graded-permittivity
sphere with respect to Eqs.~(\ref{eq:potential_field_E}),
(\ref{eq:potential_in_u}), and (\ref{eq:potential_in_v})]
over the volume of the graded-permittivity sphere
\begin{equation}
\label{eq:dipole}
    \vec{p} = \int_{\Omega}\vec{P}\diff V
        \approx \frac{\varepsilon_{\mathrm{eff}}-\varepsilon_{0}}
        {\varepsilon_{\mathrm{eff}}+2\varepsilon_{0}}
        4\pi\varepsilon_{0}r_{0}^{3}\vec{E}_{\mathrm{in}},
\end{equation}
associated to the effective permittivity $\varepsilon_{\mathrm{eff}}$ of the
graded-permittivity sphere in the quasi-static limitation.
The calculated absolute value of the dipole moment is depicted in Fig.~2(b).
Since the scattering cross section in Eq.~(\ref{eq:sca}) can be described as
\begin{equation}
    C_{\mathrm{sca}} \approx \frac{128\pi^{5}r_{0}^{6}}{3\lambda^{4}}
        \left| \frac{\varepsilon_{\mathrm{eff}}-\varepsilon_{0}}
        {\varepsilon_{\mathrm{eff}}+2\varepsilon_{0}} \right|^{2},
\end{equation}
the dipole moment in Eq.~(\ref{eq:dipole}) is proportional to the square root of the
scattering cross section.
It is shown that the variation of the dipole moment as a function
of the frequency in Fig.~2(b) is coincident with the variation of the
scattering efficiency of the scattering cross section in Fig.~2(a) for both
the graded-permittivity sphere and the normal sphere.

Besides the scattering efficiency, the scattering diagram provides detailed
information about the scattering of the electromagnetic wave in terms of
the distribution of the scattering electromagnetic field.
Figure~3 plots the numerically simulated scattering diagram
with the amplitude of the scattering electric field in both
$x$-$z$ plane and $y$-$z$ plane for both the graded-permittivity
sphere [Fig.~3(a)] and the normal sphere [Fig.~3(b)]
at different frequencies.
In general, with respect to the same incident electromagnetic wave
$\vec{E}_{\mathrm{in}}=1\,\mathrm{V}/\mathrm{m}$, the scattering electric
field by the graded-permittivity sphere is about one order of magnitude
lower than the normal sphere.
In addition, the scattering diagram of the graded-permittivity sphere performs
ultra-compact patterns due to the polarization cancellation that manifests
the broadband transparency.
Specifically, according to the variation of the permittivity of the graded-permittivity
sphere [Fig.~1(b)], it is clear that at the low frequency $\omega/\omega_{p}=0.34$
the graded-permittivity sphere behaves as a metallic-like sphere due to large part
of the graded-permittivity sphere possesses a negative permittivity close to the core.
Therefore, the scattering diagram represents a regular Mie scattering pattern but
much more compact than that of the normal sphere because the positive permittivity
of the graded-permittivity sphere near the sphere surface still causes the polarization
cancellation.
As the increasing of the frequency the polarization cancellation becomes more efficient
that directly leads to a highly compact scattering pattern at the frequency of $\omega/\omega_{p}=0.55$.
Finally, at the high frequency $\omega/\omega_{p}=0.82$, the polarization cancellation still impacts
on the scattering pattern, but the high frequency incident electromagnetic wave also
excites high order scattering that begins to demonstrate the scattering, thus the compact
scattering pattern is different from a regular Mie scattering pattern.
In contrast, the scattering diagram of the normal sphere is quite different and
it clearly illustrates the variation from the Rayleigh scattering pattern to
the Mie scattering pattern with the increased frequency.

The polarization cancellation can be demonstrated directly by exploring the
distribution of the polarization of the graded-permittivity sphere, as illustrated
in Fig.~4.
Here the polarization of the graded-permittivity sphere is studied at three
different frequencies that are coincident with the scattering diagrams in Fig.~3.
The theoretically calculated polarization distributions [Fig.~4(a)]
agree with the numerical simulation results [Fig.~4(b)] quite well.
Moreover, it is clear that the polarization near the surface of the graded-permittivity
sphere and the polarization near the core possess opposite direction with the
zero-permittivity position as their boundary [as shown in Fig.~1(b)].
According to the permittivity profile described in Eq.~(\ref{eq:grade-drude}),
the zero-permittivity position in the graded-permittivity sphere varies with respect
to the frequency so that the polarizations near the surface and the core will always
possess opposite directions over a wide frequency range.
Hence, the polarization cancellation effect will be satisfied at all frequencies
for realizing the broadband electromagnetic transparency.

Furthermore, the graded-permittivity sphere can be realized as a multi-shell
sphere with properly designed permittivity in each shell regarding the practical
applications.
According to the graded-permittivity profile shown in Fig.~1(b), a 10-shell
sphere is constructed with each shell possessing the homogeneous and isotropic permittivity
based on the mean value theorem as
\begin{equation}
    \varepsilon_{n}(\omega) = \frac{1}{\Delta r}\int_{r_{n-1}}^{r_{n}}
        \varepsilon_{g}(\omega,r)\diff r
        = 1 - \frac{\omega_{p}^{2}}{\omega^{2}}
        \left( C_{0} - C_{1}\frac{r_{n+1}^{2}+r_{n+1}r_{n}+r_{n}^{2}}
        {3r_{0}^{2}} \right),
\end{equation}
where $r_{n}=(n-1)r_{0}/10$ for $n=1,2,\ldots,10$ and $\Delta r=r_{n}-r_{n-1}$.
The scattering efficiency of the 10-shell sphere is examined and displayed in Fig.~2(a),
which is similar to that of the original graded-permittivity sphere except for small
perturbations that may be caused by the discontinuous permittivity between the adjacent
shells.
Additionally, the polarization of the 10-shell sphere is also studied in Fig.~4(c)
based on the numerical simulations.
Similar to the graded-permittivity sphere shown in Fig.~4(b), the polarization
of the 10-shell sphere also possesses opposite directions with respect to
different frequencies and the polarization cancellation leads to the broadband
electromagnetic transparency.

\section{Conclusions}
The broadband electromagnetic transparency is realized with a properly
designed graded-permittivity sphere with extremely low scattering cross section over
a broad frequency range, using both the generalized Mie scattering theory and the
numerical simulation in full-wave condition.
The dynamic polarization cancellation is achieved with the graded-permittivity sphere
in a wide frequency range for the demonstration of broadband electromagnetic transparency.
A multi-shell sphere with properly arranged homogeneous permittivity in each shell
is also proposed and studied to realize the broadband transparency phenomenon.
This work can be a fundamental study to achieve the broadband object cloaking,
and also will benefit many research areas in nanoscale optics and photonics
such as optical sensing and photodetection.

\section*{Acknowledgement}

This work is partially supported by the Energy Research and Development Center
at Missouri S\&T, the University of Missouri Research Board and Interdisciplinary
Intercampus Research Program.


\newpage                  %
\section*{Figure Captions}%

\textbf{FIG.~1}. (Colour online)
(a) Schematic diagram of the scattering of an incident plane
electromagnetic wave by a graded-permittivity sphere located
at the origin.
The electromagnetic wave is propagating in the positive $z$-direction
with the electric field in the $x$-direction and the magnetic field
in the $y$-direction.
(b) The variation of the permittivity of the well-designed
graded-permittivity sphere with respect to the frequency,
where the zero-permittivity position along the radial direction
is plotted as the intersecting curve.

\vspace{5.0mm}
\textbf{FIG.~2}. (Colour online)
(a) The variation of the efficiency factors of the scattering cross
section of the graded-permittivity sphere (black),
the normal sphere (red), and the 10-shell sphere (blue),
with respect to the frequency.
The circle markers indicate the numerical simulation results
at different frequencies.
(b) The variation of the absolute value of the dipole moments
of the graded-permittivity sphere (black curve) and the normal
sphere (red-dashed curve), which is coincident with the variation
of the efficiency factor shown in (a).

\vspace{5.0mm}
\textbf{FIG.~3}. (Colour online)
The scattering diagram of (a) the graded-permittivity sphere
and (b) the normal sphere at three different frequencies
$\omega/\omega_{p}=0.34$, $0.55$, and $0.82$,
with respect to the same incident plane electromagnetic wave
of the electric field $\vec{E}_{\mathrm{in}}=1\,\mathrm{V}/\mathrm{m}$.
The scattering diagram is represented as the distribution of
the electric field amplitude for the scattering electromagnetic
wave in the $x$-$z$ plane (red) and the $y$-$z$ plane (blue), respectively.
The electric field amplitude for the scattering electromagnetic
wave from the graded-permittivity sphere is about one order of
magnitude less than that from the normal sphere at all frequencies.

\vspace{5.0mm}
\textbf{FIG.~4}. (Colour online)
The polarization of the graded-permittivity sphere with respect
to the incident plane electromagnetic wave based on (a) the
theoretical analysis and (b) the numerical simulation,
as well as (c) the polarization of the 10-shell sphere
based on the numerical simulation at the same three different
frequencies in Fig.~3.
The polarizations of the graded-permittivity sphere always
possess opposite directions at all frequencies, leading to
the polarization cancellation that results in the broadband
electromagnetic transparency.

\newpage
\begin{figure}[htb]
    \centering
    \includegraphics[width=75mm]{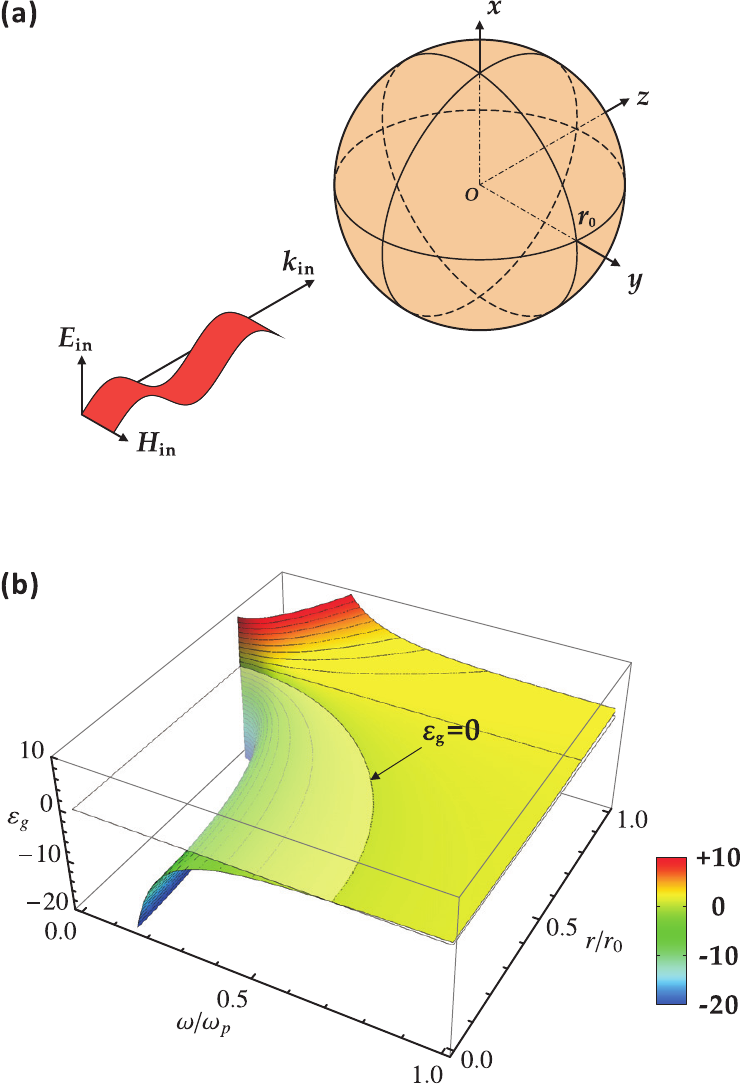}
    \caption{Lei~Sun, Jie~Gao, and Xiaodong~Yang}
\end{figure}

\newpage
\begin{figure}[htb]
    \centering
    \includegraphics[width=65mm]{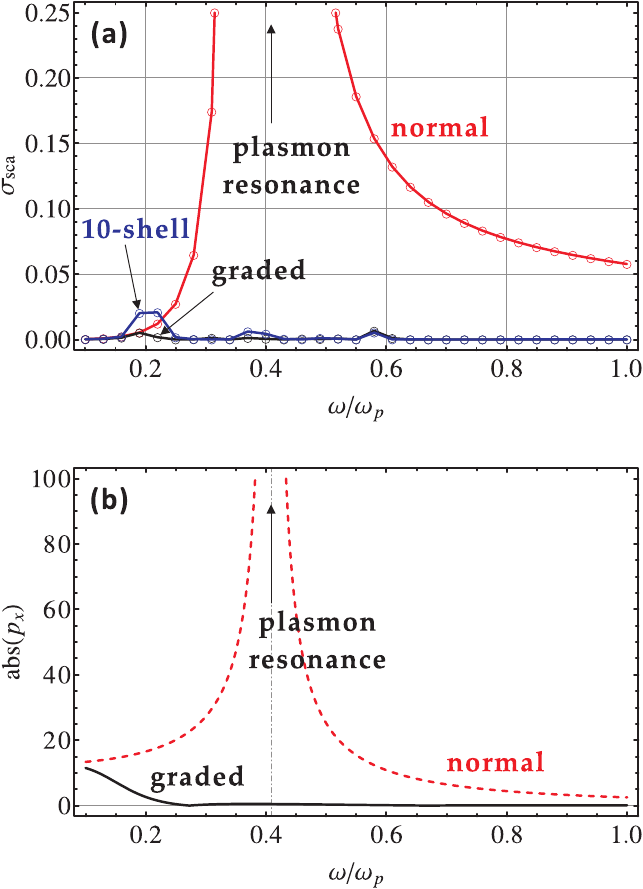}
    \caption{Lei~Sun, Jie~Gao, and Xiaodong~Yang}
\end{figure}

\newpage
\begin{figure}[htb]
    \centering
    \includegraphics[width=140mm]{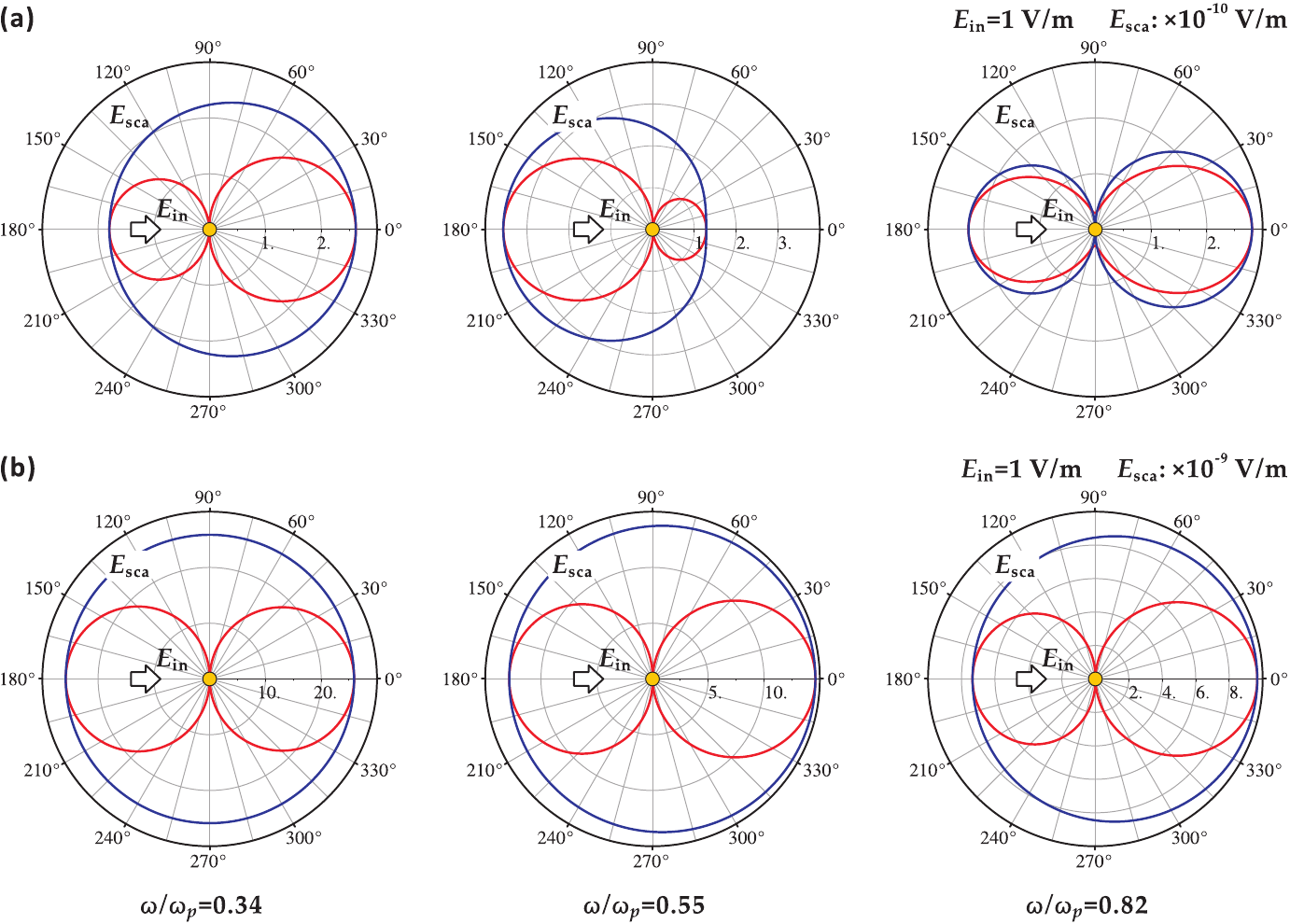}
    \caption{Lei~Sun, Jie~Gao, and Xiaodong~Yang}
\end{figure}

\newpage
\begin{figure}[htb]
    \centering
    \includegraphics[width=135mm]{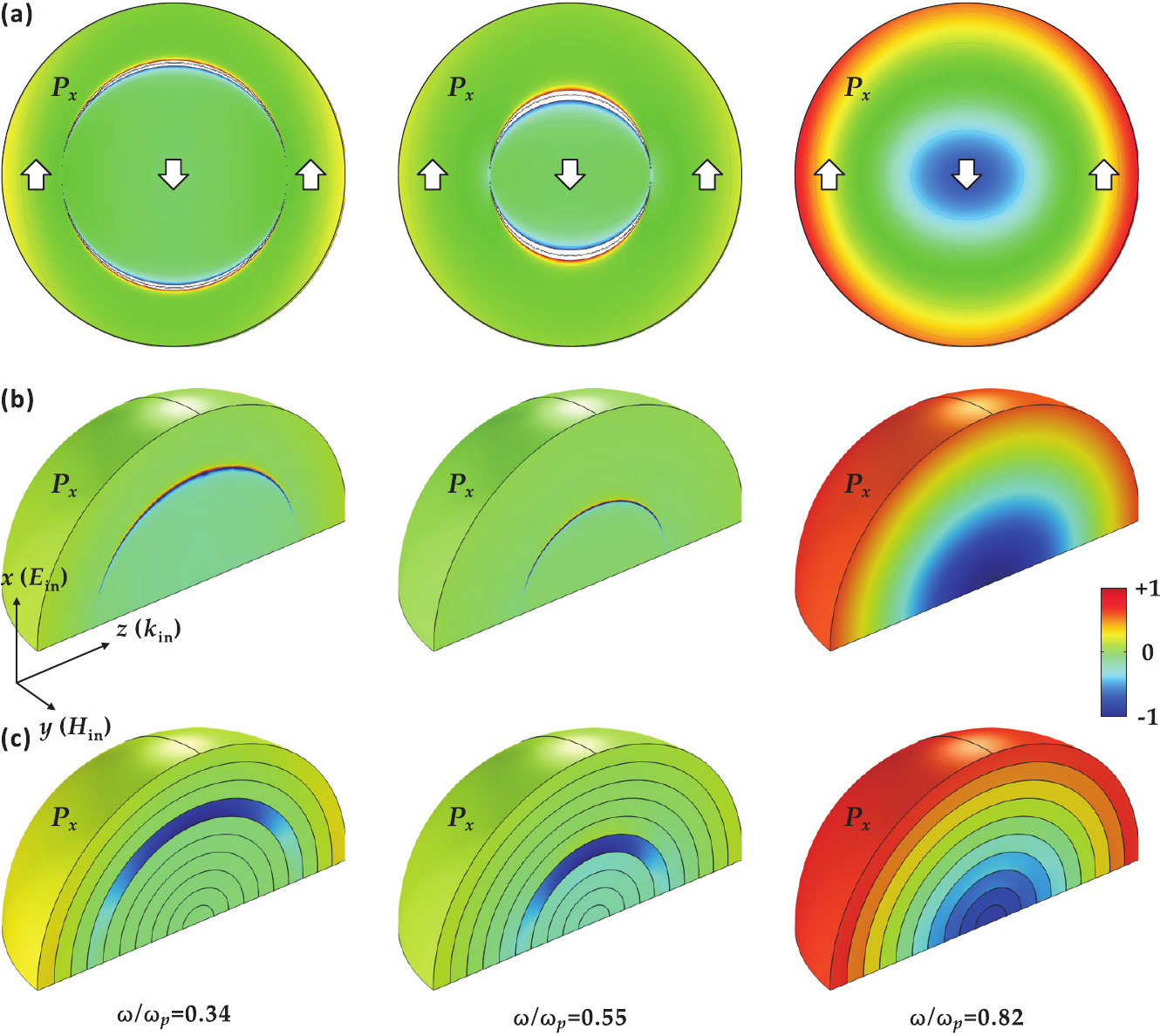}
    \caption{Lei~Sun, Jie~Gao, and Xiaodong~Yang}
\end{figure}

\end{document}